# Ion acceleration during internal magnetic reconnection events in TST-2

H. Hoshika[a], H. Zushi[b], M. Aramasu[c], H. Idei[b], A. Iyomasa[b], A. Ejiri[c], S. Ohara[d], H. Kasahara[c], Y. Kamada[c], S. Kawasaki[b], M. Sakamoto[b], K. Sasaki[a], K. Sato[b], S. Shiraiwa[c], Y.Takagi[d], Y. Takase[c], H. Nakashima[b], K. Nakamura[b], M. Hasegawa[b], K. Hanada[b], A. Higashijima[b] and T. Yamada[d]

[a] Interdisciplinary Graduate School of Engineering Sciences. Kyushu University,
[b] Research Institute for Applied Mechanics ,
[c] Graduate School of Frontier Sciences. The University of Tokyo,
[d] Department of Science. The University of Tokyo,

**Abstract**

Characteristics of ion acceleration in the internal magnetic reconnection events (IRE) have been studied by means of a neutral particle energy analyzer (NPA) in Tokyo Spherical Tokamak (TST-2). The major and minor radii are 0.38 m and 0.25m, respectively. The magnetic field strength is 0.3T and the maximum plasma current is up to 140 kA. The electron and ion temperatures are 0.4 – 0.5 keV and 0.1 keV, respectively and the electron density is ~1x10$^{19}$ m$^{-3}$. The NPA can be scanned toroidally from θ = 74° (cw) to θ = 114° (ccw), where θ = 90° corresponds to the perpendicular sightline. The direction of the plasma current is cw. The NPA signals are digitized at every 50 μs. The NPA is calibrated in the energy range of 0.1 keV < E < 8.4 keV.

When the IRE occurs, it is observed that the plasma current increases by ~ 20% and the loop voltage drops from 0.6 V to – 5 V for ~ 0.1 ms. The enhanced charge exchange flux is observed by more than one order of magnitude at ~ 1 keV for this reconnection phase. The ion temperature increases by 80 eV at IREs. The angle θ dependence of increment of Ti shows that ΔTi (θ = 74°) is higher than that for θ = 114°. This observation suggests that an ion is accelerated initially in the direction of magnetic field lines. The time evolution of the ion distribution function is simulated with a Fokker–Planck code taking into account the electric field effects.

## 1. Introduction

An internal reconnection event (IRE)[1] is one of the interesting MHD phenomenon in ST plasmas. In ref. [1] characteristics of the IRE are a rapid current rise, an increase in the soft X-ray signal at the peripheral chords, visible light spikes, and MHD activities. The current rise can be as much as 40% of the peak current. From the precursor oscillations with m=1, where m is the poloidal mode number, it is explained that a flattening of the current profile leads to large current increases.

One of the interesting aspects which are considered to be related with astrophysical phenomena, such as solar flare or solar wind, is plasma heating during the magnetic filed reconnection. Actually in STs strong ion heating has been observed [2]. In MAST the formation of suprathermal deuterium and hydrogen populations even in Ohmic plasmas following IRE was observed by a neutral particle energy analyzer. For the hydrogen beam injection at 40 keV into a deuterium plasma, similar tail component of deuterium ions are also observed and it has been explained that one mechanism is that of large angle scattering process.

Recently Helander [3] proposed that high-energy tail appeared in the ion distribution function following IREs is clearly associated with reconnection and it can be explained as a manifestation of runaway acceleration in the parallel electric filed associated with reconnection.

In a spherical tokamak TST-2 [4], ion heating has been also observed in Ohmic plasma. Although several types of MHD events are found, IREs are categorized as "strong MHD events" having a clear positive current spike. A marked phenomenon during a strong MHD event is found that the impurity CV temperature increases ~570 eV significantly during the events and then decays to the level (~98 eV) before the event. They used a visible spectrometer equipped with 11 channels photomultiplier and the Doppler broadening of CV line was measured with 10 μs resolution time. This measurement was done along the perpendicular viewing chord to the magnetic filed lines and bulk hydrogen ion temperature was not measured. Before event the relaxation time between impurity ions and hydrogen ions is ~ 10 μs, but during and after that it increases 30 ~ 250 μs depending on the temperature rise and density drop. Although the released amount of the poloidal magnetic filed energy is not inconsistent with the increment of the ion energy, heating mechanism and bulk hydrogen behavior have been left to be resolved.

Recently the neutral particle energy analyzer NPA was installed to TST-2 and the ion distribution function was measured during the "strong MHD events". In order to obtain the ion distribution function in phase space the NPA can be scanned with respect to the magnetic file line. In this paper the experimental results of the NPA associated IREs and numerical results of time dependent Fokker-Planck simulation taking into account the electric field induced by IREs will be presented.

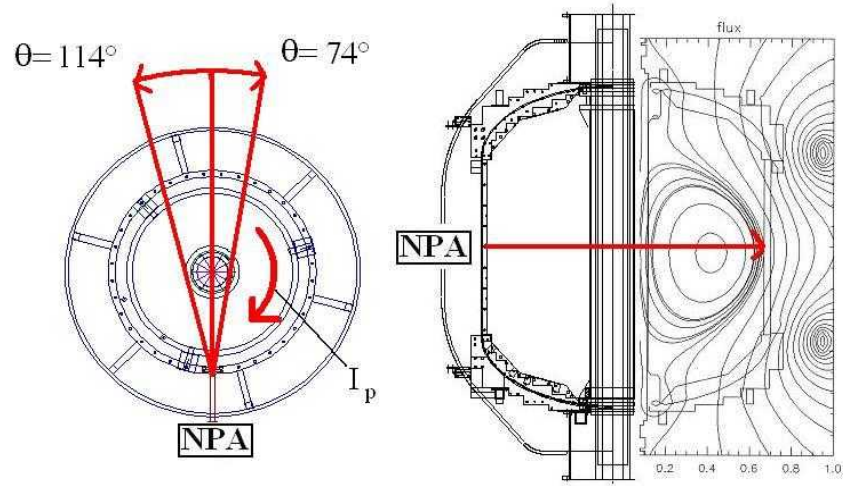

Fig.1 Top and cross sectional views of TST-2 and NPA line of sights.

## 2. Experimental apparatus

TST-2 [5] is the spherical tokamak whose parameters are as follows: major radius R~ 0.38 m, minor radius a ~ 0.25 m, aspect ratio A ~ 1.6, elongation $\kappa \leq 1.8$, toroidal magnetic filed Bt ~ 0.3 T, plasma current~ 0.14 MA and discharge duration $t_{pulse}$ ~ 0.04 s. IREs in TST2 are usually observed, particularly during the current decay phase in Ohmic phase. Observations of the current positive spikes and negative loop voltage spikes suggest a flattening of the current density profile. The line averaged density decreases and the $H_\alpha$ emission increases abruptly during an IRE, indicating a significant particle loss [4, 5].

Figure 1 shows the top view and cross section of the TST-2 and arrangement of NPA. The typical magnetic surfaces are also shown. Visible light measurement and soft X-ray measurement are done. In addition to these four sets of soft X-ray arrays are installed to study the MHD activity of IREs.

NPA is used to detect neutral particles coming from plasma via charge exchange process between hydrogen atom and proton. This system consists of drift tube, stripping gas cell, parallel electro static plates and detectors. The pressure in the analyzer is kept below $1 \times 10^{-5}$ torr by a turbo molecular pump. The ceratron is used as a detector and is operated in pulse count mode. Signals are digitized by a camac scaler at every 50 - 200 μs. The energy range is from 0.3 keV to 2.1 keV. The energy spectrum can be obtained on shot-to-shot basis assuming the discharge reproducibility. The viewing chord is 2 cm above the midplane. The NPA can be scanned with respect to the toroidal magnetic filed from 74° (clockwise direction) to 114°(counterclockwise direction). Here the angles are defined as that determined at the magnetic axis. The tangent radii are 0.19 m and 0.12m, respectively. Here the plasma current flows in the clockwise direction. The former detects co-moving ions and the latter ctr-moving ions. Since the passive NPA measurement is a line integration of the CX flux originated at each location along the line of sight, the pitch angle is not determined uniquely. With assumptions of $n_i(r) = n_i(0)(1-(r/a)^2)^{0.5}$ and $n_0(r) = n_0(0)(r/a)^2$ the most dominant pitch angle of the detected CX flux at E=0.9 keV is calculated. $n_i(0)$ is assumed $1\times10^{19}$ m$^{-3}$. Two maxima correspond to 77° (near position) and 67° (far position) for the clockwise chord of 74° and the minimum pitch angle is 56°. Four maxima exist along the counterclockwise direction of 114°, however, and they correspond to 109°(near position), 123° (inboard region), 205°(far inboard region over the tangent point) and 245°(far outboard region), respectively. The fractions of the flux normalized at the near position are 0.4, 0.2, 0.06, respectively.

## 3. Experimental results

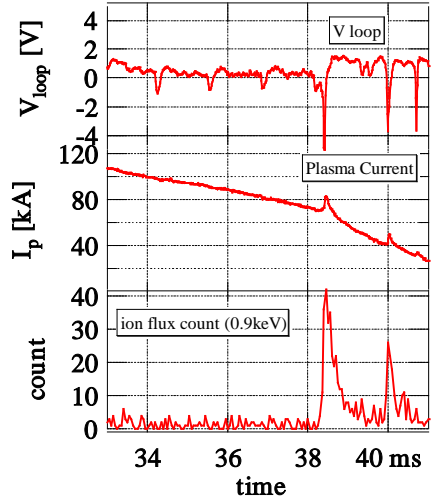

**Fig.2** Loop voltage, plasma current, and CX flux at 0.9 keV with IREs.

IREs in TST-2 are shown in Fig. 2. The multi-channel soft X-ray arrays are used to follow IRE. The precursor oscillations with an m=1/n=1 mode are found to be associated with some instability triggering IRE. The position of the safety factor q=1 surface is deduced at r/a~ 0.5 from the SX measurement and is consistent with the equilibrium calculations. The current profile broadening is suggested by positive current and negative loop voltage spikes. At IREs the bursts of NPA signal are also observed. The different point from the MAST results [3] is that this data is taken along the line of sight perpendicular to the magnetic filed lines. At t=38.1 ms Ip rises 20% and $V_{loop}$ drops by -5 V. Although the NPA flux corresponds to $H_\alpha$ bursts, the enhancement factor is quite large for NPA compared with a factor of two enhancement of $H_\alpha$. Thus it is considered that ions are heated during IREs.

Ion temperature evolution is shown in Fig.3 for three $\theta_{pitch}$. $T_i$ is ~ 100 eV before IREs and is comparable to the CV temperature [4]. When IRE occurs, $T_i$ along the line of $\theta_{pitch} = 74°$ reaches 180 eV. In order to avoid statistical variation the CX counts are summed for 0.2 ms and then $T_i$ is determined. The temperature rise $\Delta T_i$ is plotted as a function of $\theta_{pitch}$ obtained in several reproducible shots, as shown in Fig. 4. Scattering of $\Delta T_i$ is ascribed to variations in IREs ($\Delta Ip/Ip$ or $V_{loop}$). $\Delta T_i$ is highest for $\theta_{pitch}=74°$ and it is surprised that $\Delta T_i$ for $\theta_{pitch}=90°$ and 114° show positive. This is a quite different point from the MAST results [3]. For the perpendicular line of sight they measured no temperature rise during IREs.

In Fig. 5 the increment in CX flux is plotted as a function of expanded time during and after IRE. Here t=0 ms is defined as the time at which the CX flux starts to increase. For comparison $V_{loop}$ is also shown as a monitor of IRE. The decay time of the NPA flux at E=0.6 keV is ~ 0.5 ms, which is roughly equal to the ion slowing down time.

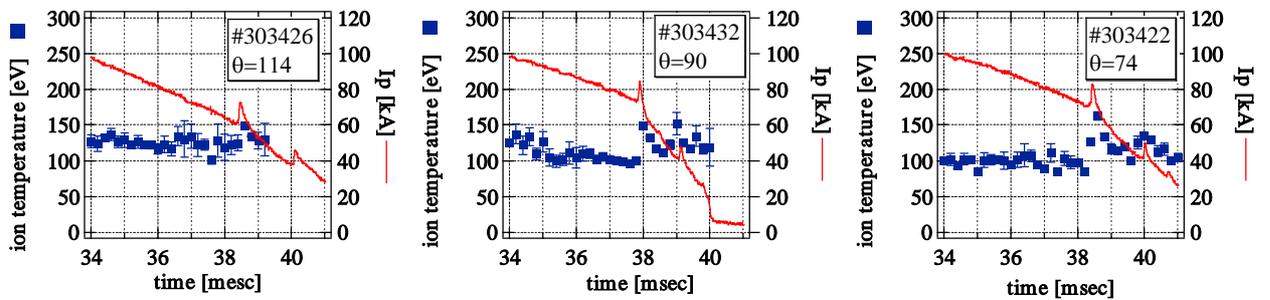

**Fig.3** Ion temperature evolution obtained along different three lines of sight.

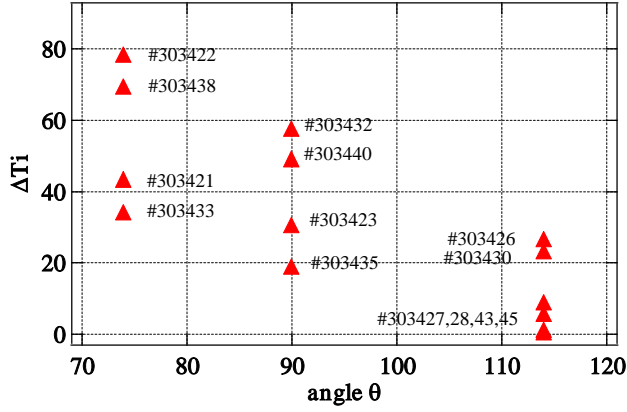
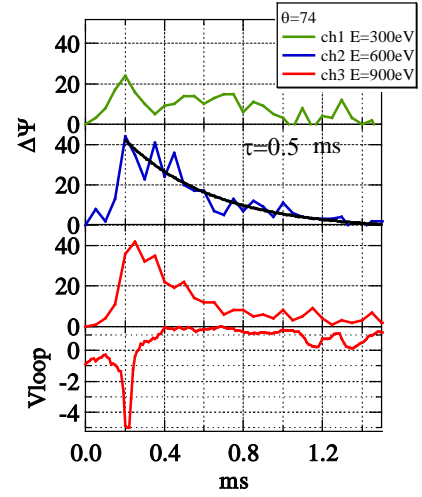

**Fig.4** The increment in Ti vs. the chord pitch angle at the magnetic axis.

**Fig.5** The CX fluxes at E=0.3, 0.6, and 0.9 keV shown in the expanded time scale.
Vloop is also plotted for the monitor of IRE.

## 4. Simulation results

Furth and Rutherford have reported in ref. [6] that ion runaway occurs when the toroidal electric filed satisfies a certain condition $E^* > E_c(3m_e/2\pi M_i)^{1/3}$, where $E^*$ is the effective electric field, $E_c$ the Dreicer field, $m_e$ the electron mass and $M_i$ mass of bulk ions, respectively. This condition is satisfied during the IRE in TST-2. The electric field is calculated ~ 45 V/m from the equilibrium code with measured magnetic probe signals and an assumed pressure profile. In order to analyze the ion acceleration a bounce averaged FP code [7] is used. In this code electrons, hydrogen ions and oxygen impurity are dealt and their temperatures are assumed 550 eV, 100 eV and 100 eV, respectively. Initial distribution functions are Maxwellian. The time period during which $E_{//}$ of 45 V/m is produced along the $v_{//}$ axis for 0.15 ms is assumed and $f(v_{//}, v\perp,t)$ is followed during that period and after that. Figure 6 shows $f(v_{//}, v\perp)$ at t=0.15 ms, just end of acceleration. Dashed, dot-dashed and dotted lines are numerical ones and correspond to the measured results taken at $\theta_{pitch}$ =74°, 90°, and 110°, respectively. Here the initial $f(v_{//}, v\perp)$ at t=0 ms, denoted by solid line, is chosen to fit the measured one before the IRE. Figure 7 shows the contour of $f(v_{//}, v\perp)$ at t=0.15 ms. Thus simulations show a clear shift of the center and ion heating is found along $\theta_{pitch}$=74° and ion cooling along $\theta_{pitch}$=114°. The result for $\theta_{pitch}$=90° remained unchanged.

The experimental results show significant heating. It should be noted that the NPA signal is the line integrated one, therefore along the sight of line ions in the acceleration region and in the deceleration region are recorded. For $\theta_{pitch}$=74° the NPA can view ions having $\theta_{pitch}$=74° only if they are on the magnetic axis. The spatial effects should be considered to compare observations with numerical results. For $\theta_{pitch}$=90°, however, the line of sight is always perpendicular to the magnetic field line. Therefore observed ion heating is not consistent with the numerical result. According to the loop voltage analysis done in MAST [3], the electric filed is much larger in the center of the plasma than at the edge, and the sign of it is in the parallel to Ip at the center where the current drops, but it is reversed at the inboard side where the current rises. Since NPA signals along $\theta_{pitch}$= 114° detect ions near the inboard edge, where the negative voltage is expected to be induced and ions can be accelerated in the anti-parallel to Ip. If this effect dominates ions accelerating in the parallel direction, the apparent accelerated spectrum is also possible. $H_\alpha$ measurement shows the enhancement at the inboard edge, therefore the enhanced cx flux is possibly expected along this chord. For $\theta_{pitch}$=74° this is not true. Another possible explanation is a mechanism of acceleration perpendicular to the magnetic filed line. Taking into account the spacial structure of reconnection, induced electric filed and the NPA line of sight will be left in future.

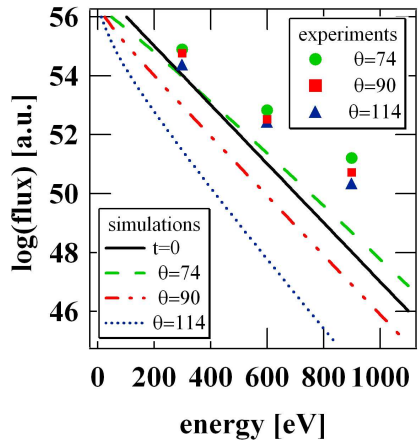
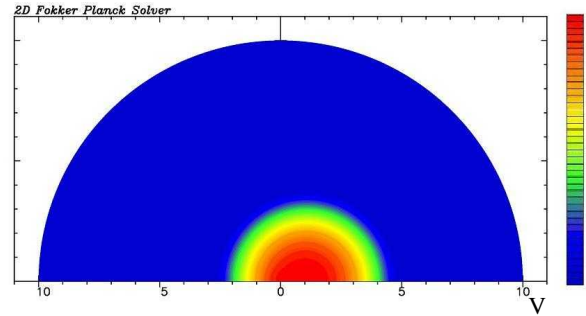

**Fig.6** The numerical results of energy spectra at t= 0.15 ms along the three pitch angles. Black is the initial spectrum, and green, red, blue lines correspond to $\theta_{pitch}$= 74°, 90°, and 114°, respectively. Circles, squared, and triangles are also experimental data for $\theta_{pitch}$= 74°, 90°, and 114°, respectively.

**Fig.7** The contour plot of f(v//, v⊥) at t=0.15 ms. v is normalized the thermal velocity.

## 5. Discussion and Summary

The CX flux measurements have been performed in TST-2 spherical tokamak. It shows bursts corresponding to IRE. This aspect is similar to that observed in MAST. Ion temperature is increased from ~ 100eV to 180 eV for co- moving ions, which is accelerating direction. Increment in $T_i$ is sustained for ~ 0.5 ms, which is much longer than the slowing down time and rather order of energy confinement time. The equilibrium calculation gives the toroidal electric filed of 45 V/m, which satisfies the ion runaway condition. The numerical simulation by the time dependent bounce averaged Fokker-Planck code shows that the ion acceleration is possible even for 0.15 ms, which is the order of IRE duration.

There are two unresolved problems; namely ion heating perpendicular to the magnetic filed and anti-parallel to the magnetic filed. This is not expected by the toroidal electric induced during reconnection and related ion runaway mechanism. Although the pitch angle scattering time is ~ 0.2 ms at 0.7 keV ions, a clear sequential acceleration is not observed in the CX fluxes.  Large angel scattering rather than multiple small angle scattering or the excitation of the waves heating ions during the IREs are possible candidates [3]. For the anti-parallel acceleration, it should be considered that the line of sight and the negative electric filed at the inboard region contribute apparent acceleration along this chord. The critical electric field is evaluated 0.7 V/m at the center, which accelerates ions in the co-direction. Although the negative electric filed is much low and is induced near the edge, this field also satisfies the ion runaway condition if $n_e/T_e$ near the edge is the same as the center. If this is true, accelerated ions can be detected along this line of sight. Coupled the equilibrium calculation with 2D Fokker Planck calculations and numerical integration of f(v) along the line of sight are underway.

## Acknowledgements

The authors thank Prof. M. Azumi at JAERI for providing us his Fokker-Planck code, which helped us a better understanding of ion distribution function during IREs. This work has been partially performed under the framework of joint-use research in RIAM Kyushu University and the bi-directional collaboration organized by NIFS. This work was partly supported by the Sasakawa Scientific Research Grant.